\documentclass[
    amsmath,
    amssymb,
    ]{rspublic}
\usepackage{graphicx}
\usepackage{subfigure}
\usepackage{bbm}
\usepackage{calc}
\usepackage{cancel}
\usepackage{hyperref}
\usepackage{type1cm}
\usepackage{type1ec}
\usepackage[utf8]{inputenc}
\usepackage{placeins}

\usepackage{listings}
\usepackage{color}
\definecolor{darkgreen}{rgb}{.1,.4,.15}
\definecolor{darkred}{rgb}{.5,.1,.2}
\definecolor{darkblue}{rgb}{.2,.3,.6}
\newcommand{\cdim}[1]{{\rm cdim} #1 }
\newcommand{\SLH}[1][{}]{\left(\mathbf{S}_{#1}, \mathbf{L}_{#1}, H_{#1}\right)}
\newcommand{\ket}[1]{\left | #1 \right\rangle}
\newcommand{\bra}[1]{\left \langle #1 \right|}
\newcommand{\expect}[1]{\left\langle #1 \right\rangle}

\usepackage[numbers]{natbib}
\title[Specification of photonic circuits using QHDL]{Specification of photonic circuits using Quantum Hardware Description Language}
\author[N. Tezak, A. Niederberger, D. S. Pavlichin, G. Sarma, and H. Mabuchi]{Nikolas Tezak$^1$, Armand Niederberger$^{1,2}$, Dmitri S. Pavlichin$^1$, Gopal Sarma$^1$, and Hideo Mabuchi$^1$} 
\affiliation{${}^1$ Edward L.\ Ginzton Laboratory, Stanford University, Stanford, CA 94305, USA \\
${}^2$ SUPA, Department of Physics, University of Strathclyde, Glasgow G4 0NG, UK
}

\date{\today}
\begin{document}
\maketitle
\begin{abstract}
    {Open systems, quantum statistical methods, optical computers}
Following the simple observation that the interconnection of a set of quantum optical input-output devices can be specified using structural mode VHSIC Hardware Description Language (VHDL), we demonstrate a computer-aided schematic capture workflow for modeling and simulating multi-component photonic circuits. We describe an algorithm for parsing circuit descriptions to derive quantum equations of motion, illustrate our approach using simple examples based on linear and cavity-nonlinear optical components, and demonstrate a computational approach to hierarchical model reduction.
\end{abstract}

\noindent Ongoing advances in materials science, nanoscale synthesis and lithography are establishing key technical requirements for the fabrication of complex nanophotonic circuits.
Critical applications for such circuitry can be foreseen in ultra-low power photonic signal processing and interconnects, sensor networks, and quantum information technology.
Although current research in nanophotonics focuses largely on the physics of individual devices such as resonators and waveguides, there are substantial new challenges to be addressed in the analysis and design of circuits comprising large-scale networks of interconnected components with dynamic optical nonlinearities.
For premier applications, quantum-optical network models will be required in order to capture fluctuations, coherence, and entanglement effects that may be decisive drivers of the overall circuit performance.
Suitable theoretical frameworks exist but require cumbersome algebraic manipulations for the derivation of multi-component models, pointing to the need for computer-aided paradigms for generating quantum-optical equations of motion from high-level circuit representations that can be manipulated more intuitively by circuit engineers.
Simply put, we are rapidly approaching a new phase of research in quantum nonlinear photonics in which we will need user-friendly circuit design tools like the ones we have long exploited in classical electronics.

In this paper we propose and demonstrate a modeling and simulation workflow based on schematic capture using a Quantum Hardware Description Language (QHDL) for nanophotonic circuits, which we will define as a proper subset of the standard VHSIC Hardware Description Language (VHDL). Our approach utilizes a mixture of common open-source software packages and custom processing scripts to provide a high-level, modular interface to the quantum circuit `algebra' of Gough and James~\cite{Gough2008,Gough2009} (which generalizes earlier work on cascaded open quantum systems by Carmichael~\cite{Carmichael1993} and by Gardiner~\cite{Gardiner1993}). The natural hierarchical organization of VHDL and the schematic capture workflow should facilitate future work on model reduction and design abstractions for nanophotonic circuits, which seems essential given the extremely high dimension (variable count) associated with many-component quantum models.

In the following sections we first review the formal setting of $\SLH$ component models and the concatenation and series and products as introduced by Gough and James, which have recently been used to derive quantum nonlinear photonic circuit models by hand or using custom-coded computer algebra scripts~\cite{Kerckhoff2010,Kerckhoff2011,Mabuchi2011a,Mabuchi2011}. While we will restrict our attention here to linear and cavity nonlinear optics, it should be noted that the $\SLH$ formalism can in principle be used to describe hybrid circuits incorporating suitable spintronic, nanomechanical~\cite{Jacobs2010}, and/or quantum-electronic components. Likewise, the approach we describe here could be extended straightforwardly to admit static Bogoliubov components as described in~\cite{Gough2010}. We then review the proposed syntax of QHDL and illustrate its use in the specification of a simple interferometer as a network of elementary optical components. After describing methods that can be used to parse QHDL circuit descriptions to derive quantum equations of motion for analysis and/or numerical simulation, we illustrate the full schematic capture workflow using an example of constructing a bistable latch from cavity nonlinear optical components. The paper closes with a brief consideration of model reduction in the $\SLH$ context.

\section{Modeling quantum circuitry}

Within this section we use $\{Q_j,\, j = 1, 2, 3, \dots, N\}$ to denote individual quantum input-output components. We clearly distinguish between input and output ports and do not consider bi-directional ports, although for physical reasons every input port is assumed to have an associated output port and {\it vice versa}.

\subsection{The circuit algebra}
\label{sub:subsection_name}

Our modeling workflow is based on the Gough-James synthesis results for open quantum systems~\cite{Gough2008, Gough2009}, which provide a purely algebraic method to derive quantum Markov models for a network of interconnected quantum components.

A component with an equal number $n$ of input and output channels is described by the parameter triplet $\left(\mathbf{S}, \mathbf{L}, H\right)$, where $H$ is the effective internal \emph{Hamilton operator} for the system, $\mathbf{L} = (L_1, L_2, \dots, L_n)^T$ the \emph{coupling vector} and $\mathbf{S} = (S_{jk})_{j,k=1}^n$ is the \emph{scattering matrix}, whose elements are themselves operators.

Each element $L_k$ of the coupling vector is given by an operator that describes the system's coupling to the $k$-th input channel. Similarly, the elements $S_{jk}$ of the scattering matrix are given by system operators describing the scattering between different field channels $j$ and $k$. The only mathematical conditions on the parameters are that the Hamiltonian operator be self-adjoint and the scattering matrix be unitary:
\begin{align}
    H^\dagger = H \text{ and } \mathbf{S}^\dagger \mathbf{S} = \mathbf{S} \mathbf{S}^\dagger = \mathbbm{1}_n.\nonumber
\end{align}

The master equation~\cite{Gardiner2000} corresponding to a given $\left(\mathbf{S}, \mathbf{L}, H\right)$ model is
\begin{align}
    \frac{d \rho_t}{dt} & = -i[H, \rho_t] + \sum_{j=1}^n\left( L_j\rho_t L_j^* - \frac{1}{2}\left\{ L_j^*L_j, \rho_t \right\}\right)\label{eq:ME}
\end{align}
Here $[A,B]\equiv AB - BA$ and $\{A,B\}\equiv AB + BA$, while $\rho_t$ is a density matrix describing the evolving state of the internal degrees of freedom. It is also straightforward to obtain the quantum filtering equations~\cite{Bouten2007,Wiseman2010} for stochastic simulation of a given $\left(\mathbf{S}, \mathbf{L}, H\right)$ model.

While the scattering matrix elements $S_{jk}$ do not appear in Eq.~(\ref{eq:ME}) they are required for the composition rules described below, which can be used to derive the overall parameter triplet for a network of interconnected quantum input-output components. The $\left(\mathbf{S}, \mathbf{L}, H\right)$ circuit algebra plus simple correspondences such as Eq.~(\ref{eq:ME}) provide all that is needed to obtain overall equations of motion for complex photonic circuits.

\begin{figure}[htb]
    \centering
    \subfigure[$Q_1 \boxplus Q_2$]{
        \label{fig:concatenation}
        \raisebox{.1cm}{\includegraphics[width=2.5cm]{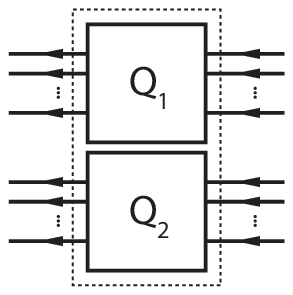}}}
    \subfigure[$Q_2 \lhd Q_1$]{
        \label{fig:series}
        \raisebox{.3cm}{\includegraphics[width=3.5cm]{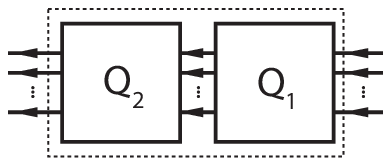}}}
    \subfigure[{$[Q]_{k \to l}$}]{
        \label{fig:feedback}
        \includegraphics[width=2.5cm]{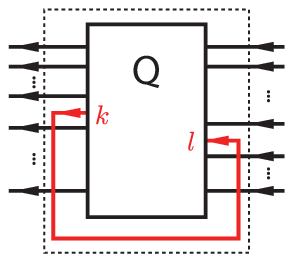}}
    \caption{\label{fig:circuit_operations} Basic operations of the Gough-James circuit algebra.}
\end{figure}

In \cite{Gough2009}, Gough and James have introduced two operations that allow for the construction of arbitrary networks of optical \emph{feedforward} circuits:
\begin{enumerate}
    \item The \emph{concatenation} product (cf.~Figure \ref{fig:concatenation}) describes a formal adjoining of two systems in which there is no optical scattering between the systems:
\begin{align}
    \left(\mathbf{S}_1, \mathbf{L}_1, H_1\right) \boxplus \left(\mathbf{S}_2, \mathbf{L}_2, H_2\right)
      \equiv \left(\begin{pmatrix} \mathbf{S}_1 & 0 \\ 0 & \mathbf{S}_2 \end{pmatrix}, \begin{pmatrix}\mathbf{L}_1 \\ \mathbf{L}_1 \end{pmatrix}, H_1 + H_2 \right)
\end{align}
Note however, that even without optical scattering, the two subsystems may interact via shared quantum degrees of freedom. A simple example of this scenario is given by a two-mode resonator (such as a ring-resonator) with an atom that interacts with both optical modes, but in which there is no direct scattering between the modes.
\item The \emph{series} product (cf.~Figure \ref{fig:series}) describes a configuration in which two systems $Q_j = \left(\mathbf{S}_j, \mathbf{L}_j, H_j \right)$, $j=1,2$ possessing an equal number of channels $n$ are connected in such a manner that all output channels of $Q_1$ are fed into the corresponding input channels of $Q_2$. The resulting system is then given by
\begin{align}
\left(\mathbf{S}_2, \mathbf{L}_2, H_2 \right) \lhd \left( \mathbf{S}_1, \mathbf{L}_1, H_1 \right)
\equiv \left(\mathbf{S}_2 \mathbf{S}_1,\mathbf{L}_2 + \mathbf{S}_2\mathbf{L}_1 , H_1 + H_2 + \Im\left\{\mathbf{L}_2^\dagger \mathbf{S}_2 \mathbf{L}_1\right\}\right),
\end{align}
where we define the imaginary part of an operator as $\Im\{A\} \equiv {A - A^\dagger \over 2i}$.
\end{enumerate}
To make the network operations complete, one additional rule is required: The \emph{feedback} operation (cf.~Figure \ref{fig:feedback}) describes the case where the $k$-th output channel of a system with $n\ge 2$ channels is fed back into the $l$-th input channel. The result is a system with $n-1$ channels:
\begin{align}
    \left[\left(\mathbf{S}, \mathbf{L}, H \right)\right]_{k \to l} \equiv \left(\tilde{\mathbf{S}}, \tilde{\mathbf{L}}, \tilde{H}\right)
\end{align}
Formulae for the resulting parameter triplet are provided in \ref{sec:feedback_parameters}.

Note that the series product can be expressed in terms of the concatenation and feedback operations ({\it e.g.}, for two components with $n=1$ we have $Q_2\lhd Q_1=[Q_1\boxplus Q_2]_{1\to 2}$), and consequently, the latter two operations are sufficient to perform all network calculations.  However, the series product is a useful shorthand and allows for a more intuitive network expression.

For use in the following we define the identity system with $n$ channels
\begin{align}
    \mathbbm{1}_n \equiv \left(\mathbf{1}_n, \mathbf{0}, 0 \right),
\end{align}
where $\mathbf{1}_n = (\delta_{kl})_{k,l = 1}^n$ is the identity matrix in $n$ dimensions,
as well as the channel permuting system
\begin{align}
    P_\sigma \equiv \left(\mathbf{P}_\sigma, \mathbf{0}, 0 \right),
\end{align}
where the permutation matrix is defined by $\mathbf{P}_\sigma \equiv \left(\delta_{k,\sigma(l)}\right)_{k,l=1}^n$.
This definition ensures that $P_{\sigma_2} \lhd P_{\sigma_1} = P_{\sigma_2 \circ \sigma_1}$.
\subsection{The QHDL syntax}
\label{sec:qhdl}

QHDL is a subset of structural VHDL \cite{Pedroni2004}, which we will use as a formal syntax for specifying photonic circuits in terms of interconnections among referenced quantum input-output components. These components can themselves represent composite networks of subcomponents, facilitating hierarchical approaches to photonic circuit design. It is useful to start with a set of basic components such as beamsplitters and phase-shifts, as well as linear and non-linear cavity models with one or more coupling mirrors\footnote{It is important to note that the Gough-James circuit algebra cannot be used to build dynamical systems from static components, {\it e.g.}, it cannot create the Fock space and operator algebra for an optical resonator mode as an automatic result of cascading beamsplitters and phase-shifts in the configuration of a ring cavity. All such dynamic components therefore must be implemented as primitive $\left(\mathbf{S}, \mathbf{L}, H\right)$ models.}, which can be collected in a shared library file. The set of such \emph{primitive} components within a QHDL software environment can of course be extended at any time.

Within the context of a single QHDL file, the exact physical model (parameter triplet) of any referenced component is left unspecified except for its external ports and parametric dependencies. This approach allows the circuit designer to operate at a high level of abstraction, facilitating last-minute substitution of alternative physical component models (including effective models with reduced simulation complexity) into a given interconnection topology.

In the following section, we will introduce the QHDL syntax by means of a very simple circuit that realizes a Mach-Zehnder interferometer.

\begin{figure}[htb]
    \centering
        \fbox{\includegraphics[width=5cm]{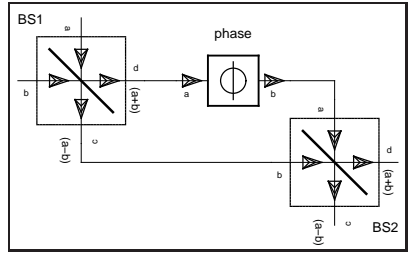}}
    \caption{\label{fig:mach_zehnder} A basic Mach-Zehnder setup.}
\end{figure}

A QHDL file begins with the \emph{entity} declaration, which defines the abstract interface of the circuit being specified: it specifies a list of named input and output ports (of the overall circuit), which are required in order for the circuit itself to be callable as a composite QHDL component, as well as any numeric parameters required for physical modeling. Note that we require that all input ports appear before all output ports.
\begin{lstlisting}[linerange={1-5}, caption={Entity declaration}]
entity Mach_Zehnder is
    generic (phi_mz: real := 0);
    port    (In1, VacIn: in fieldmode; Out1, Out2: out fieldmode);
end Mach_Zehnder;
\end{lstlisting}

For this entity we must then have one or more \emph{architecture} declarations in the same QHDL file. These provide alternative ways of realizing the internal structure of the circuit. The architecture declaration consists of a head which specifies the interfaces of all \emph{components} used in the architecture body and all internal \emph{signals}. The component declarations are very similar to the entity declaration- they serve to establish an interface for each subcomponent.
\begin{lstlisting}[linerange={6-16}, caption={Architecture head}]
architecture structure_MZ of Mach_Zehnder is
    component beamsplitter
        port (a, b: in fieldmode; c, d: out fieldmode);
    end component beamsplitter;
    
    component phase
        generic (phi: real);
        port (a: in fieldmode; b: out fieldmode);
    end component phase;    
    
    signal bs1_phase, bs1_bs2, phase_bs2: fieldmode;
\end{lstlisting}

The architecture body then consists of a series of \emph{instance} assignments for each occurrence of any of the previously specified component types. Each instance assignment specifies the relationship between the component-instance parameters and the entity parameters.  In addition, it specifies a port map detailing how the component-instance is connected to the internal signals or the external ports.
\begin{lstlisting}[linerange={17-33}, caption={Architecture body}]
begin
    BS1: beamsplitter
        port map (a => In1, b => VacIn, c => bs1_bs2, d => bs1_phase);
    phase: phase
        generic map (phi => phi_mz);
        port map (a => bs1_phase, b => phase_bs2);
    BS2: beamsplitter
        port map (a => phase_bs2, b => bs1_bs2, c => Out1, d => Out2);
end structure_MZ;

\end{lstlisting}

In the port map, each internal component port is assigned to either an entity port or a signal.
Any instance \emph{in} (\emph{out}) port must be connected either to an entity \emph{in} (\emph{out}) port or to a signal that is connected to another instance's \emph{out} (\emph{in}) port.
\begin{lstlisting}[linerange={24-24}, caption={Port map statement}]
        port map (a => phase_bs2, b => bs1_bs2, c => Out1, d => Out2);
\end{lstlisting}
Each signal therefore connects exactly two ports: one instance input and one instance output \emph{or} one instance input (output) and an entity input (output).

\subsection{Parsing a network} 
\label{sub:parsing_a_network}
Here we present a simple algorithm to parse a general network into a circuit expression. We assume that the QHDL file has been preprocessed such that we have the lists of ports, components, instances, signals and port mappings in native data structures accessible to our algorithm.
\begin{enumerate}
    \item We denote the list of internal signals by $S$. For each instance assignment $j=1,2\dots N$ in the architecture body:
    \begin{itemize}
        \item Generate the network triplet $Q_j = (\mathbf{S}_j,\mathbf{L}_j,H_j)$ with the correct parametrization as specified in the \texttt{generic map} statement.
        \item Generate the correctly ordered\footnote{As defined via the the component declaration in the architecture head.} list of input port names $I_j$ and the correctly ordered list of output port names $O_j$ where each portname is entry is of the form \emph{instance-name:port-name}.
    \end{itemize}
    \item Concatenate all triplets $Q = Q_1 \boxplus Q_2 \boxplus \dots \boxplus Q_N$ and similarly concatenate the input and output port lists $I = I_1 + I_2 + \dots + I_N$ and $O = O_1 + O_2 + \dots + O_N$
    \item For each internal signal $s \in S$ concatenate the full circuit triplet $Q$ with a single channel identity system $\mathbbm{1}_1$ resulting in $Q_f^{(0)} = Q \boxplus \mathbbm{1}_{|S|}$,
    \item Now, each element in the full list of output ports $O$ corresponds to an entry of the form \emph{instance-name:port-name}. Make copies of $O' = O$ and $S' = S$ and iterate over all output ports in the following fashion:

        If the output port is connected to a global output (i.e. an entity output port), continue to the next entry. \\
        If the output port is connected to the $j$-th signal in the \emph{current} signal list $S'$, let $k$ be the index of the output port in the \emph{current} output port $O'$ list and update the model triplet $Q_f^{(n)} \to Q_f^{(n+1)} =  [Q_f^{(n)}]_{k \to M + j}$, where $M = |O'|$ is the length of the current output port list. Then, remove the $k$-th entry from $O'$, and the $j$-th entry of $S'$.
    \item Now, let $M_f = |S'|$ and iterate over a copy of the input port list $I' = I$ and a new copy of the signal list $S'' = S$:

    If the input port is connected to a global input (i.e. an entity input port), continue to the next entry. \\
    If the input port is connected to the $j$-th entry of $S''$ update $Q_f^{(n)} \to Q_f^{(n+1)} =  [Q_f^{(n)}]_{M_f + j \to k}$ where $k$ is the index of the current port in $I'$. Then, remove the $k$-th entry of $I'$ and the $j$-th entry of $S''$.

    \item By construction, the only remaining ports of our resulting triplet $Q_f^{\rm res}$ lead to global/entity ports.
    Iterating over $O'$ and the list of entity output ports $O_E$, construct a suitable permutation $\sigma_{\rm out}$ that maps every output port index from $O'$ to the correct index of the entity output port.
    In a similar fashion, iterate over $I'$ and the list of entity input ports $I_E$ to generate a permutation $\sigma_{\rm in}^{-1}$, mapping the indices from $I'$ to the correct indices of the entity input ports within $I_E$. Then, invert this permutation $\sigma_{\rm in} = (\sigma_{\rm in}^{-1})^{-1}$ to obtain a mapping from $I_E$ to $I'$. Finally, the model triplet for the circuit is given by
    \begin{align*}
        Q_{\rm final} = P_{\sigma_{\rm out}} \lhd Q_f^{\rm res} \lhd P_{\sigma_{\rm in}}.
    \end{align*}
\end{enumerate}

If one is interested in working with the actual network expressions as opposed to the more concrete level of the actual Hilbert space operators, there exist other, more complex, approaches to parsing a network, which directly yield simpler overall network expression. Combined with a sufficiently sophisticated set of circuit expression simplification rules, the above algorithm works just as well.

\subsection{The QHDL workflow} 
\label{sub:the_qhdl_workflow}

The circuit design workflow relies heavily on symbolic computer algebra methods. Using symbolic algebra, rather than working with numerical matrix representations of all the operators appearing in the component parameter triplets, makes it possible to view the overall circuit $\SLH$ in analytic form. It also allows the designer to defer choosing the values of numerical parameters, which could be convenient for optimization scenarios, as well as details such as the upper photon-number limits to use for truncated Fock spaces in numerical simulations.

In fact we can define our own algebraic types, operations and simplification rules not just for Hilbert space operators and scalar coefficients, but also for circuit algebra components. This approach enables us to extend the hierarchical design principle even to our compiled QHDL component library, as will become clear in the following outline of the modeling workflow:
\begin{enumerate}
\item {\bf Circuit design} In step 1, we visually compose the circuit using a schematic capture tool and then export to QHDL\footnote{In our case we have modified the VHDL exporting functionality of the gEDA toolsuite to generate QHDL.} or directly describe the circuit in text-based QHDL. Since QHDL describes the connections between \emph{functional} entities, it is not necessary at this stage to specify how referenced components are implemented.
\item {\bf Component model specification} The QHDL file is then parsed to generate the circuit expression in which referenced components appear as symbols.  This expression is stored in a library file along with information about model parameters and the component names of the referenced subcomponents. Note that a library file can be treated as a standalone entity for future circuit designs.  When this file is imported at runtime, the referenced subcomponent models are dynamically loaded from their respective library files. Now, the full $\SLH$ parameter triplet can be generated by explicitly evaluating the circuit algebra operations. By means of the symbolic operator algebra, the final operator matrices and the Hamiltonian are still in fully symbolic form, which can be used to generate the quantum master equation or an appropriate stochastic differential equation in symbolic form. This allows for the application of analytical model reduction techniques before turning to purely numerical methods.
\item {\bf Numerical simulation} Define all scalar model parameters and (truncated) Hilbert space dimensions, and compute the behavior of the circuit.
\end{enumerate}
In Table~\ref{tab:software_stack} we list the necessary software tools to implement the QHDL circuit design workflow. We plan to publicly release our custom tools in the near future.

\section{An example of the QHDL workflow} 
\label{sec:examples}

In this section, we present a detailed example applying the QHDL workflow to the design, analysis, and simulation of an all-optical $\overline{SR}$-latch as recently proposed in~\cite{Mabuchi2011, Mabuchi2011a}.
The elementary component models $\{(\mathbf{S}_j,\mathbf{L}_j,H_j)\}$ required for this circuit are the following:
\begin{enumerate}
\item {\bf Beamsplitters}
    \begin{align*}
    Q_{BS}(\theta) \equiv \left(\begin{pmatrix} \cos \theta & -\sin \theta \\ \sin \theta & \cos \theta \end{pmatrix}, \mathbf{0}, 0\right),
    \end{align*}
\item {\bf Phase-delays} $U(\phi) \equiv (e^{i\phi},0,0)$
\item {\bf Coherent displacements} $W(\alpha) \equiv (1, \alpha, 0)$ which models a laser source outputting a coherent field with amplitude $\alpha \in \mathbbm{C}$,
\item {\bf Kerr-nonlinear cavity} (here a unidirectional ring cavity with two input/output ports)
    \begin{align*}
        Q_{K}(\Delta, \chi, \{\kappa_j\}) = \left(\mathbbm{1}_2, \begin{pmatrix} \sqrt{\kappa_1 }a \\ \sqrt{\kappa_2} a\end{pmatrix}, \Delta a^\dagger a + \chi a^\dagger a^\dagger a a \right)
    \end{align*}
\end{enumerate}
As the circuits we discuss here are meant to be used as logical gates in larger circuits, we need not include the laser sources in our circuit schematics. Instead, they can easily be added at the level of the circuit algebra by feeding a concatenated block of laser displacements (sources) into the full network
$Q_{\text{with input}} = Q \lhd (W_{\alpha1} \boxplus W_{\alpha2} \boxplus \dots \boxplus W_{\alpha n})$.

\begin{table}
    \centering
    \caption{\label{tab:software_stack} List of software components necessary to realize our QHDL-workflow.}
\scriptsize{\begin{tabular}{ p{3.8cm}@{\hspace{.1cm}}|@{\hspace{.1cm}} p{3.8cm} @{\hspace{.1cm}}|@{\hspace{.1cm}} p{3.8cm} }
 \textbf{Requirement} & \textbf{Our solution} &  \textbf{Alternatives}\\
\hline
\hline
Graphical \textbf{schematic capture} tool with VHDL/QHDL export capabilities & gschem and  gnetlist from the gEDA suite \cite{gEDA} &  Graphical design tool from system modeling environments / modeling languages, such as Modelica \cite{Modelica} \\
\hline
\textbf{QHDL-Parser} that computes the circuit expression & A custom parser written in Python using the open source PLY \cite{PLY} package & A parser for computing a circuit expression from the Modelica specification, written in, e.g. \emph{Mathematica} \\
\hline
\textbf{Symbolic computer algebra system} with support for: non-commutative operator algebra, commutative scalar algebra coefficient, operator-valued matrix algebra and the Gough-James circuit algebra & A custom computer algebra system written in Python \cite{Python} and interfacing with SymPy \cite{SymPy} for the scalar coefficient algebra & \emph{Mathematica} \cite{Mathematica} plus an implementation of the Gough-James circuit algebra \\
\hline
\textbf{Numerical backend} to convert symbolic operator expressions into matrices and simulate the system & Custom algorithms for solving the Master equation as well as quantum stochastic differential equations implemented in Python and C using optimized numerical libraries for linear algebra \cite{ScipyNumpy,MKL}.& The Quantum Optics Toolbox \cite{QOToolbox} for MATLAB \cite{MATLAB} or similar library for modeling the dynamics of open quantum systems, such as QuTIP \cite{Johansson}
\end{tabular}}
\end{table}

\subsection{The two-cavity pseudo-NAND-latch} 
\label{sub:a_simplified_latch}
\begin{figure}[htb]
    \centering
    \includegraphics[width=12cm]{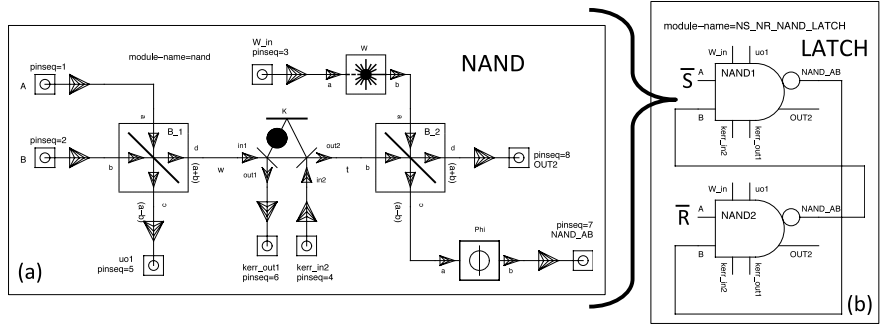}
    \caption{\label{fig:pseudo_nand_latch_circuits} Pseudo-NAND circuit schematic (a) as created with gschem and its device symbol embedded as a component in a SR-NAND-latch circuit (b).}
\end{figure}

\begin{figure}[htb]
    \centering
    \subfigure[Circuit expression \eqref{eq:pseudo_nand_unsimplified} as generated by the QHDL-Parser]{
        \label{fig:pseudo_nand_expression_unsimplified}\fbox{\includegraphics[width=8cm]{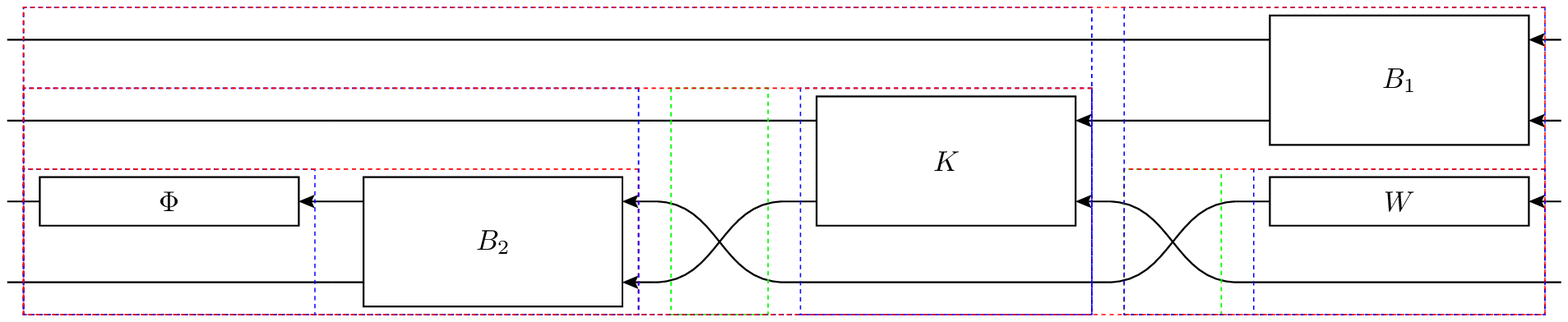}}}

    \subfigure[Simplified expression \eqref{eq:pseudo_nand_simplified}]{
        \label{fig:pseudo_nand_expression_simplified}\fbox{\includegraphics[width=8cm]{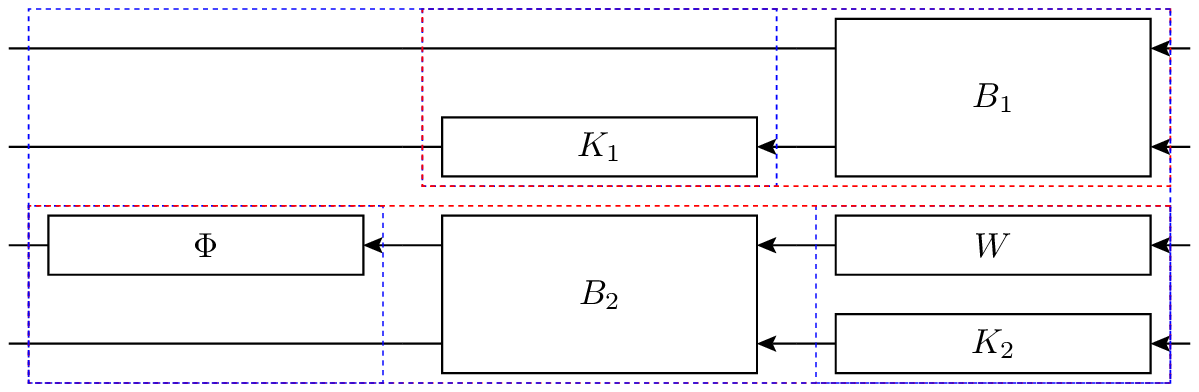}}}
    \caption{\label{fig:pseudo_nand_expressions} Pseudo-NAND circuit expression visualizations. As can easily be verified visually, the simplified expression follows from decomposing $K = K_1 \boxplus K_2$ and `pulling' $K_2$ down into the fourth row. These expression simplifications are automatically performed by our symbolic circuit algebra software.}
\end{figure}

We have recently proposed~\cite{Mabuchi2011} several different optical circuits to realize three classical logic gates: an AND-gate, a NOT-gate with integrated fanout of two and a combined (but imperfect) NAND-gate (Figure~\ref{fig:pseudo_nand_latch_circuits}(a)), which in the following we will call pseudo-NAND gate as it only works properly when at any given time at least one input is in the `on' state. The first two of these gates used in sequence also realizes a NAND gate, but the advantage of the pseudo-NAND is that it requires only a single Kerr-nonlinear cavity component.
The QHDL workflow can be readily applied to design the pseudo-NAND circuit and automatically generate the circuit expression in terms of its components\footnote{Here we represent a permutation $\sigma = \begin{pmatrix} 1 & 2 & \dots & n \\ \sigma(1) & \sigma(2) & \dots & \sigma(n)\end{pmatrix}$ just by its image tuple $\left(\sigma(1) \, \sigma(2) \, \dots \, \sigma(n)\right)$.}:
\begin{align}\label{eq:pseudo_nand_unsimplified}
   &\left( \mathbbm{1}_1 \boxplus \left( \left( \mathbbm{1}_1 \boxplus \left( \left( {{\Phi}} \boxplus \mathbbm{1}_1 \right) \lhd {B_2} \right) \right) \lhd P_{(1\,3\,2)} \lhd \left( {K} \boxplus \mathbbm{1}_1 \right) \right) \right) \nonumber \\
& \qquad \lhd \left( {B_1} \boxplus \left( P_{(2\,1)} \lhd \left( W \boxplus \mathbbm{1}_1 \right) \right) \right)
\end{align}
where the beamsplitter symbols are defined by $B_1 \equiv Q_{BS}({\pi \over 4})$, $B_2 \equiv Q_{BS}(\theta)$, the output correction phase is $\Phi \equiv U(\phi)$, the constant coherent displacement component is $W \equiv W(\beta)$ and the Kerr cavity is given by $K \equiv Q_K(\Delta, \chi, \kappa_1 = \kappa_2 = \kappa)$.
The network expression \eqref{eq:pseudo_nand_unsimplified} looks complicated but can be verified easily by comparing its visual representation (Fig.~\ref{fig:pseudo_nand_expression_unsimplified})\footnote{These visualizations were automatically generated using another software tool we have implemented.} with the original circuit schematic.
Moreover, since the scattering matrix of $K$ is in block-diagonal form, it is possible to decompose the cavity component $K = K_1 \boxplus K_2$, where the Hamiltonian of $K$ can be assigned to either of the two blocks. Upon substituting this decomposable form into the expression, the automatic expression simplification built into our circuit algebra implementation yields the following form:
\begin{align} \label{eq:pseudo_nand_simplified}
    \Big\{ \left( \mathbbm{1}_1 \boxplus K_1 \right) \lhd B_1 \Big\} \boxplus \Big\{ \left( \Phi \boxplus \mathbbm{1}_1 \right) \lhd B_2 \lhd \left(W \boxplus K_2 \right) \Big\},
\end{align}
which is visually represented in Figure~\ref{fig:pseudo_nand_expression_simplified}.
The numerical model parameters as given in \cite{Mabuchi2011} are $\theta = 0.891$, $\chi = -5/6$, $\Delta = 50$, $\kappa = 25$, $\phi = 2.546$ and the auxiliary constant input amplitude is given by $\beta = -34.289 - 11.909i$. The coherent input amplitudes corresponding to the logical signals `on' and `off' are then given by $\alpha = 22.6274$ and 0, respectively.

As in classical circuit theory, two NAND-gates in a mutual feedback configuration, as shown in Figure~\ref{fig:pseudo_nand_latch_circuits}(b), can be used to realize a latch with inverted inputs $\overline{\rm S}$ and $\overline{\rm R}$.
A latch features controllable bistable behavior and thus realizes a single-bit memory unit. It has two inputs: S(ET) and R(ESET), which can be activated individually to control the internal logical state to `on' or `off', respectively. Ideally, when both S and R are `off' (HOLD-condition), the internal state remains stable. In practice, quantum fluctuations and noisy inputs lead to spontaneous switching between the two internal states. One of the design goals is thus to decrease the rate at which this spontaneous switching occurs.
The QHDL code as produced by gnetlist \cite{gEDA} (slightly edited to be more concise) can be found in Listing~\ref{lst:pseudo_nand_latch} and the circuit component library file generated by the QHDL-parser is presented in Listing~\ref{lst:pseudo_nand_latch_qos} in~\ref{sec:latch_library}.
\begin{lstlisting}[caption={QHDL source for the pseudo-NAND latch.}, label={lst:pseudo_nand_latch}, float=t]
--pseudo-NAND latch
ENTITY NS_NR_NAND_LATCH IS
    PORT (NS, W1, kerr2_extra, NR, W2, kerr1_extra : in fieldmode;
          BS1_1_out, kerr1_out2, OUT2_2, BS1_2_out, kerr2_out2, OUT2_1 : out fieldmode);
END NS_NR_NAND_LATCH;

ARCHITECTURE netlist OF NS_NR_NAND_LATCH IS
    COMPONENT nand
        PORT (A, B, W_in, kerr_in2 : in fieldmode; 
              uo1, kerr_out1, NAND_AB, OUT2 : out fieldmode);
    END COMPONENT;    

    SIGNAL FB12, FB21 : fieldmode;      -- feedback signals

BEGIN
    NAND2 : nand
    PORT MAP (
        A => NR, B => FB12, W_in => W2, kerr_in2 => kerr2_extra,
        uo1 => BS1_2_out, kerr_out1 => kerr2_out2, NAND_AB => FB21, OUT2 => OUT2_2);

    NAND1 : nand
    PORT MAP (
        A => NS, B => FB21, W_in => W1, kerr_in2 => kerr1_extra,
        uo1 => BS1_2_out, kerr_out1 => kerr1_out2, NAND_AB => FB12, OUT2 => OUT2_1);
END netlist;
\end{lstlisting}
Substituting the individual component models into the circuit expression yields the full triplet $\SLH[0]$ for the latch. Finally, after feeding in the coherent input signals $\overline{\rm S}$ and $\overline{\rm R}$ into their respective ports:
$\SLH = \SLH[0] \lhd (W(\overline{\rm S}) \boxplus \mathbbm{1}_2 \boxplus W(\overline{\rm R}) \boxplus \mathbbm{1}_2)$ the parameters assume the following form
\begin{align}
    \mathbf{S} & =
       \left( \begin{array}{c|c}
           \mathbf{S}_1 & \mathbf{0} \\
           \hline
           \mathbf{0} & \mathbf{S}_2
           \end{array} \right),
           \mbox{ where } \mathbf{S}_1 = \mathbf{S}_2  = \begin{pmatrix}{1 \over \sqrt{2}} & -{\cos \theta e^{i \phi} \over \sqrt{2}}  &  {\sin \theta  e^{i \phi}  \over \sqrt{2}} \\
              {1 \over \sqrt{2}} & {\cos \theta e^{i \phi} \over \sqrt{2}} & -{ \sin \theta  e^{i \phi} \over \sqrt{2}} \\
              0 & \sin \theta & \cos \theta \end{pmatrix},
\end{align}
\begin{align}
 \mathbf{L} & =   \begin{pmatrix}  \sqrt{{\kappa \over 2}}\sin \theta  e^{i \phi}  \;  b  + {\overline{\rm S} \over \sqrt{2}} - {\beta \over \sqrt{2}}  \cos \theta e^{i \phi}  \\
        \sqrt{\kappa}   \;  a  - \sqrt{{\kappa \over 2}}\sin \theta  e^{i \phi}  \;  b  + {\overline{\rm S} \over \sqrt{2}}  + {\beta \over \sqrt{2}}  \cos \theta e^{i \phi}\\
         \sqrt{\kappa} \cos \theta  \;  b   +  \beta \sin \theta  \\
         \sqrt{{\kappa \over 2}}\sin \theta  e^{i \phi} \; a   + {\overline{\rm R}\over \sqrt{2}}  - {\beta \over \sqrt{2}} \cos \theta e^{i \phi} \\
        \sqrt{\kappa} \;   b - \sqrt{{\kappa \over 2}}\sin \theta  e^{i \phi} \;  a   +  {\overline{\rm R} \over \sqrt{2}} +  {\beta \over \sqrt{2}} \cos \theta e^{i \phi}\\
         \sqrt{\kappa} \cos \theta  \;  a  + \beta \sin \theta
         \end{pmatrix},
\end{align}
\begin{align}
    H
        & = \Delta  \left(a^\dagger a +  b^\dagger b\right) + \chi  \left(a^\dagger a^\dagger a a +  b^\dagger b^\dagger b b\right)   - {\kappa  \over \sqrt{2}}   \sin \theta \sin \phi  \left( a b^\dagger + a^\dagger b \right) \\ \nonumber
        &+ {\sqrt{2\kappa}  \over 4}i \left[  \left(  \overline{\rm S}^* +  \beta^* \cos \theta e^{- i \phi}\right)  a  -
                                            \rm{h.c.} \right]
  + {\sqrt{2\kappa}  \over 4}i \left[ \left ( \overline{\rm R}^* +  \beta^* \cos \theta e^{- i \phi}\right)  b -
                                            \rm{h.c.} \right].
\end{align}
Due to the symmetry of the underlying circuit model, the model parameters are invariant under exchange of the two pseudo-NAND gates, which corresponds to simultaneously exchanging $\overline{\rm S} \leftrightarrow \overline{\rm R}$, $(a,a^\dagger) \leftrightarrow (b, b^\dagger)$,  $(L_1,L_2,L_3) \leftrightarrow (L_4, L_5, L_6)$ and $\mathbf{S}_1 \leftrightarrow \mathbf{S}_2$.
This symmetry suggests that the most likely candidates for the internal logical states `on' and `off' correspond to the case where one internal cavity mode is in a high power state and the other one in a low power state and the opposite case, obtained by exchanging the cavities states. This is indeed the case, and in fact it follows from the basic way in which we have designed our pseudo-NAND gate;
`on' $\Leftrightarrow$ \{NAND1 cavity power is low, NAND2 cavity power is high\} and `off' $\Leftrightarrow$ (NAND1 cavity power is high, NAND2 cavity power is low).

To understand our model's dynamic behavior we turn to numerical methods. The simulation of this model is carried out by representing the operators as numerical matrices in a truncated product basis of Fock-states of total dimension $N^2=75^2 = 5625$\footnote{{\it I.e.}, each individual cavity basis is given by $\{\ket{0},\ket{1},\dots \ket{N-1}\}$.}. We carried out a large number of quantum jump trajectory simulations~\cite{QOToolbox,Haroche2006}
with the following sequence of alternating input conditions: 0.5 time units of SET, 5 units of HOLD, 0.5 units of RESET, 5 units of HOLD (repeated twice).
\begin{figure}[htb]
    \centering
    \includegraphics[width=10cm]{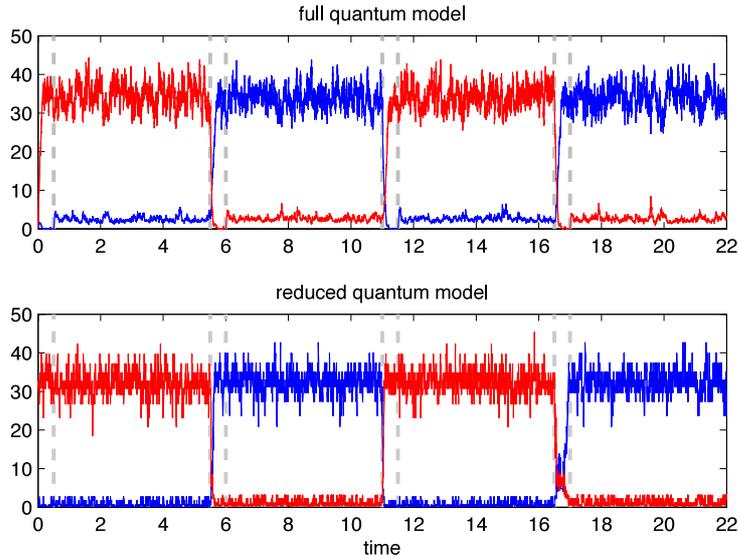}
    \caption{\label{fig:latch_traces} Simulated input sequence for the full pseudo-NAND latch model (upper) and our reduced model (lower). The red trace (lighter, in grayscale) is given by $\expect{a^\dagger a}$ and the blue trace presents $\expect{b^\dagger b}$. The SET and RESET input conditions, marked by the short intervals between dashed vertical lines, induce transitions to their respective target latch states: `on' corresponds to the $a$-mode being in a high photon number state, while `off' corresponds to a high photon number in mode $b$.}
\end{figure}
The upper plot in Figure~\ref{fig:latch_traces} presents a typical simulated trace where the system is subjected to this sequence of input conditions. We generally find that the SET and RESET input conditions successfully drive the system into the desired cavity states, while the cavities remain in their states during the HOLD condition. Although a simulation of the full master equation
is feasible using current HPC hardware and sparse matrix storage~\cite{Mabuchi2011}, quantum jump simulations exhibit the inherently bistable nature of our synthesized latch more clearly.

\subsection{Model reduction in the $\SLH$ context} 
\label{sub:MR}

As the latch could readily be used as a component in more complex circuits, such as flip-flops or even quantum memories~\cite{Kerckhoff2010,Kerckhoff2011}, it would be highly desirable to reduce the Hilbert space dimension $N^2$ required to represent it. Since we are working with quantum circuit models, we are ultimately limited by exponential scaling of the state space with the number of components (although it may be possible to develop efficient simulation procedures when components are only weakly entangled, as should be the case in ultra-low power classical signal processing). However, there is clearly much to be gained by developing accurate model reduction procedures that allow us to replace high-dimensional {\it ab initio} models for components within a circuit by much lower-dimensional effective models. Such model-reduction strategies could presumably be applied hierarchically. As in the classical theory of signals-and-systems there are many potential strategies for dimensional reduction of quantum input-output models, although little work has yet been done on this subject~\cite{Bouten2008,Gough2010a}. Here we describe an empirical approach, similar to the classical strategy of approximating Markov chains~\cite{Kushner2001}, which utilizes numerical simulation and statistical analysis to derive an effective $\SLH$ model for the pseudo-NAND latch suitable for embedding within a more complex circuit.

Our approach is based on the assumption that the device state can be inferred with reasonable accuracy from a small number of observables. We can then construct a dynamic model just in terms of these parameters~\cite{Nielsen09}.
By generating many quantum jump trajectories for the full $\SLH$ model, we generated many time series for the expectation values of the cavity field photon numbers $\expect{a^\dagger a}$ and $\expect{b^\dagger b}$ under the three valid input conditions $\{\text{HOLD},\text{SET}, \text{RESET}\}$. We can now coarse-grain the 2-dimensional space of expectation values and associate an internal model state $i \in \{1, 2,\dots, M\}$ with each bin that is actually visited during the trajectory simulations, but due to the high correlations between the cavity photon numbers, we are actually able to obtain good results by performing this coarse-graining or `binning' procedure in terms of the single quantity
\begin{align*}
    D &\equiv \expect{a^\dagger a} - \expect{b^\dagger b},
\end{align*}
implying that within the two-dimensional configuration space our system always stays fairly close to a one-dimensional submanifold.

By analyzing the observed transitions between these reduced states for each input condition $\xi \in \{\text{HOLD},\text{SET}, \text{RESET}\}$, we calculate an empirical estimate $\hat{\mathbf{P}}^{(\xi)} = (p_{ij}^{(\xi)})_{i,j = 1}^M$ of the conditional transition probabilities $p_{ij}^{(\xi)} = P(x_{n+1} = j | x_n = i, \,\xi)$ and thus model the system in terms of a discrete time Markov chain with a set of conditional transition probabilities for each particular input condition $\xi$. The time step $\delta t$ of the discrete Markov chain corresponds to the interval at which we sampled our original continuous-time system.
We now wish to get back to a description that is compatible with our $\SLH$ formalism. In the following we briefly outline a procedure to do this:
For a temporally homogeneous Markov jump process with an even number of states $i \in \{1, 2,\dots, M \}$ and transition rate matrix\footnote{\textbf{Q} is also often referred to as the \emph{generator matrix}.} $\mathbf{Q} = (\gamma_{ij})_{i,j = 1}^M$ we can define a $K$-channel $\SLH[0]$ model with states corresponding directly to the Markov process states $\{\ket{1}, \ket{2},\dots, \ket{M}\}$ via
\begin{align}
    \mathbf{S}_0 & = \mathbf{1}_K,\\
    \mathbf{L}_0 & = (\sqrt{\gamma_{i_1j_1}} \ket{j_1}\bra{i_1}, \; \dots,  \sqrt{\gamma_{i_Kj_K}} \ket{j_K}\bra{i_K})^T \\
    H_0 & = 0,
\end{align}
where the components of $\mathbf{L}_0$ drive transitions $\{i_k \to j_k, k = 1,2 \dots K\}$ and $K$ is given by the number of positive transition rates $\gamma_{ij} > 0$. By construction, as one may verify by writing down the master equation, this system always collapses into a purely classical mixture of the coarse-grained states. Equivalently, in a quantum jump trajectory simulation, after the first quantum jump, the state is always given by a single such state. In fact, in such a trajectory simulation, this system behaves exactly like the original Markov jump process.
Neglecting for now that our original model has three different input conditions $\xi \in \{\text{HOLD},\text{SET}, \text{RESET}\}$ and thus three different conditional transition matrices $\hat{\mathbf{P}}^{(\xi)}$, we first discuss how to move from the discrete time Markov chain model to a continuous time Markov jump process.
Rephrasing this question, we can ask the following: is there a Markov jump process with conditional transition probability matrix $\mathbf{P}(t)$ that `looks like' our Markov chain when stroboscopically probed at fixed time intervals $\delta t$? If our Markov chain has transition matrix $\hat{\mathbf{P}}$, then we need to determine a generator matrix $\mathbf{Q}$ such that
\begin{align}
\hat{\mathbf{P}} = \mathbf{P}(t = \delta t) \equiv e^{\delta t\mathbf{Q}} \approx \mathbf{1} + \delta t \mathbf{Q} + O(\delta t^2).
\end{align}
If our sample interval $\delta t$ is sufficiently small, we may define
\begin{align}
    \hat{\mathbf{Q}} \equiv {1 \over \delta t} (\hat{\mathbf{P}} - \mathbf{1})
\end{align}
as an approximation to the conditional rate matrix.
We now carry out the procedure outlined above to create a model $\SLH[0]$ that realizes the HOLD condition. The transition rates of the HOLD condition alone lead to a system that has two bistable clusters of states with low state indices and high state indices, respectively.

To account for the input-controlled switching in the SET and RESET conditions, we extend our model by concatenating it with a second model that explicitly includes the input fields $\SLH[\rm \overline{SR}] = \SLH[1] \lhd (W(\overline{S}) \boxplus W(\overline{R}) \boxplus \mathbbm{1}_2)$.
Hence, in the SET and RESET conditions, the HOLD transitions continue, but we drive further transitions through this additional component.
Here, $\SLH[1]$ is given by
\begin{align}
    \mathbf{S}_1 &= \mathbf{1}_4 - \begin{pmatrix}
            \Sigma_S^\dagger \Sigma_S   & 0                             & -\Sigma_S^\dagger              & 0                 \\
            0                               &  \Sigma_R^\dagger \Sigma_R & 0                             & -\Sigma_R^\dagger  \\
            -\Sigma_S                        & 0                             &  \Sigma_S \Sigma_S^\dagger & 0                 \\
            0                               & -\Sigma_R                      & 0                             & \Sigma_R \Sigma_R^\dagger
        \end{pmatrix} \\
    \mathbf{L}_1 &= - \alpha ( 1 - \Sigma_S^\dagger \Sigma_S,\; 1 - \Sigma_R^\dagger \Sigma_R,\;  \Sigma_S,\; \Sigma_R )^T \\
    H_1 & = 0,
\end{align}
where the `drift' operators $\Sigma_S$ and $\Sigma_R$ are defined by
\begin{align*}
    \Sigma_S &\equiv \ket{M-4}\bra{M-1} + \ket{M-6}\bra{M-3} +  \dots +\ket{M/2 - 1}\bra{M/2 + 2}, \\
    \Sigma_R &\equiv \ket{5}\bra{2} + \ket{7}\bra{4} + \dots + \ket{M/2 + 2}\bra{M/2 - 1}.
\end{align*}
For our simulation we chose $M=38\ll N^2$, but the general ansatz works for a range of different $M=4k + 2$ with sufficiently large $k$.
The drift operators satisfy $\Sigma_{R/S}^2  = 0$, as well as the projection relations $(\Sigma_{R/S}^\dagger \Sigma_{R/S})^2 = \Sigma_{R/S}^\dagger \Sigma_{R/S}$, $(\Sigma_{R/S}^\dagger \Sigma_{R/S})^2 = \Sigma_{R/S}^\dagger \Sigma_{R/S}$ and $\Sigma_{R/S} \Sigma_{R/S}^\dagger \Sigma_{R/S} = \Sigma_{R/S}$.
These relations suffice to show that $\mathbf{S}_1$ as defined above is indeed unitary.
To make sense of the effect of this extension to our model, consider now what happens for the different input conditions:
In the HOLD condition $\overline{S} = \overline{R} = \alpha$ the input fields cancel out all elements of the coupling vector $\mathbf{L}_1$ and we have $\SLH[\rm \overline{SR}] = \left(\mathbf{S}_1, \mathbf{0}, 0\right)$, i.e. the transition dynamics of our full system $\SLH[\rm \overline{SR}] \boxplus \SLH[0]$ are simply given by those of $\SLH[0]$ alone.

\begin{figure}[htb]
    \centering
    \includegraphics[width=10cm]{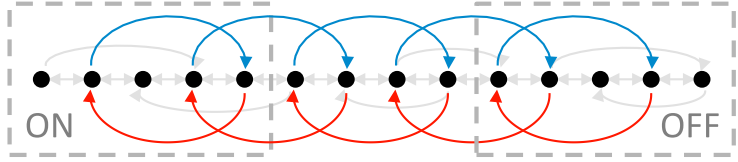}
    \caption{\label{fig:state_cartoon} Here we schematically visualize the state space and the transitions of the reduced model. The SET transitions (red, from right to left) introduce a drift that drives the system to the states on the left, corresponding to the logical `ON' state of the latch. The RESET transitions (blue, from left to right) have the opposite effect. The HOLD transitions (depicted in grey) are always active, but in the absence of additional SET and RESET transitions only very rarely lead to a switch of the logical latch state.}
\end{figure}

In the SET condition, however, we have $\overline{S} = 0, \overline{R} = \alpha$ and thus
\begin{align*}
\SLH[\rm \overline{SR}] = \left(\mathbf{S}_1, (- \alpha ( 1 - \Sigma_S^\dagger \Sigma_S),\; 0,\; - \alpha \Sigma_S, 0 )^T,0\right).
\end{align*}
The full system $\SLH[\rm \overline{SR}] \boxplus \SLH[0]$ now features the drift operator as an additional transition operator\footnote{The second non-zero element of $\mathbf{L}_{\rm \overline{SR}}$, which is a projection operator, does not affect the transition dynamics due to the fact that our system is never in a superposition of two states.} $- \alpha \Sigma_S$ which induces transitions $\{ M-1\to M-4, M-3\to M-6, \dots, M/2+2 \to M/2 -1 \}$ with constant rate $|\alpha|^2$. Together with the HOLD transitions, these lead to a drift from states with high index (corresponding to the logical `off' state of the latch) to those with low index (`on').
On the other hand, in the RESET condition, the situation is reversed. Now the other transition operator of $\mathbf{L}_1$ is canceled out and the non-zero transition operator $- \alpha \Sigma_R$ drives transitions in the reverse direction $\{ 2\to 5, 4\to 7, \dots, M/2 - 1 \to M/2 + 2 \}$, again with constant rate $|\alpha|^2$. In Figure~\ref{fig:state_cartoon} we visualize the transition structure of the model schematically.

Note also that we can emulate the state-dependent coherent output field(s) of the latch by concatenating a triplet $\SLH[\rm out]$ that re-routes bias input fields via state-dependent scattering into one or more output channels. For example we could use $\SLH[\rm \overline{SR}] \boxplus \SLH[0] \boxplus (\mathbf{S}_{\rm out},\mathbf{L}_{\rm out},0)$ where
\begin{align}
\mathbf{S}_{\rm out} = \sum_{i=1}^M \vert i\rangle\langle i\vert \begin{pmatrix} e^{i\phi_{1i}}\cos \theta_i & -e^{i\phi_{1i}}\sin \theta_i \\
e^{i\phi_{2i}}\sin \theta_i & e^{i\phi_{2i}}\cos \theta_i \end{pmatrix},\quad \mathbf{L}_{\rm out} = \mathbf{S}_{\rm out}(\beta',0)^T,
\end{align}
where $\beta'$ is the complex amplitude of a bias field and the parameters $\{\theta_i,\phi_{1i},\phi_{2i}\}$ are chosen such that the outputs of $(\mathbf{S}_{\rm out},\mathbf{L}_{\rm out},0)$ vary as desired with the internal state $\vert i\rangle$.
Having thus created a reduced model that mimics the desired input-output behavior in $\SLH$ form, we can use it to replace the full latch model in more complex circuits.
If we had already specified a QHDL file for such a circuit, we could simply replace\footnote{In principle it should be possible to include the reduced model as an alternative architecture for the latch entity and to select whether or not to use it in place of the full model at compile-time using a VHDL configuration file. However this would require some enhancements to the QHDL-Parser to correctly handle the $K$ extra (vacuum) input ports required by the reduced model to drive spontaneous transitions among the internal states.} the referenced latch component with the reduced model component. Re-parsing this modified QHDL-file would then yield a computationally more tractable model for simulations.

\section{Conclusion} 
\label{sec:conclusion}
In this paper we have described the use of QHDL, a quantum hardware description language, to facilitate the analysis, design, and simulation of complex networks constructed from interconnected quantum optical components. We have also presented a parsing algorithm for obtaining quantum equations of motion from the QHDL description. QHDL can be used as the basis for a schematic capture workflow for designing quantum circuits that automates many of the conceptually challenging and computationally demanding aspects of quantum network synthesis. As QHDL inherits the hierarchical structure of VHDL, its use may facilitate the crucial development of hierarchical model reduction methods for quantum nonlinear photonics.

Important future directions for QHDL research include simulation strategies for exploiting weak entanglement among components, stability analysis and design optimization of QHDL-based models~\cite{Niederberger}, and the incorporation of techniques from static program analysis and formal verification to assist in the design of complex, hierarchically defined photonic components. While we have emphasized classical photonic logic~\cite{Mabuchi2011} as a tutorial paradigm for QHDL in this paper, emerging ideas in quantum information processing and quantum sensing/metrology may provide even more compelling applications for QHDL as a convenient and extensible modeling framework.

\acknowledgements{
This work is supported by DARPA-MTO under Award No.\ N66001-11-1-4106, by the National Science Foundation under Grant No.\ PHY-1005386, and by the SU2P Program of Stanford University and Research Councils UK.
}

\appendix{Reduced parameters in case of signal feedback} 
\label{sec:feedback_parameters}
Upon feeding the $k$-th output channel of a system $Q = \SLH$ back into its $l$-th input, we get a system $\left[\;\SLH\;\right]_{k \to l}  = \left(\tilde{\mathbf{S}}, \tilde{\mathbf{L}}, \tilde{H}\right)$ with one less channel $\cdim \left[\;Q\;\right]_{k \to l} = \cdim Q - 1$,
where effective parameters are then given by \cite{Gough2008}
\begin{align}
    \tilde{\mathbf{S}} & = \mathbf{S}_{\cancel{[k,l]}} +  \begin{pmatrix} S_{1l} \\ S_{2l} \\ \vdots \\ S_{k-1\, l} \\ S_{k+1\, l} \\ \vdots \\ S_{n l}\end{pmatrix}(1 - S_{kl})^{-1}  \begin{pmatrix} S_{k 1} & S_{k2} & \cdots & S_{kl-1} & S_{kl+1} & \cdots & S_{k n}\end{pmatrix},
\end{align}
\begin{align}
    \tilde{\mathbf{L}} & = \mathbf{L}_{\cancel{[k]}} + \begin{pmatrix} S_{1l} \\ S_{2l} \\ \vdots \\ S_{k-1\, l} \\ S_{k+1\, l} \\ \vdots \\ S_{n l}\end{pmatrix} (1 - S_{kl})^{-1} L_k,   \quad \tilde{H}  = H + \Im\left\{\ \left[\sum_{j=1}^n L_j^\dagger S_{jl}\right] (1 - S_{kl})^{-1} L_k \right\}.
\end{align}
Here we have written $\mathbf{S}_{\cancel{[k,l]}}$ as a shorthand notation for the matrix $\mathbf{S}$ with the $k$-th row and $l$-th column removed and similarly $\mathbf{L}_{\cancel{[k]}}$ is the vector $\mathbf{L}$ with its $k$-th entry removed.
These resulting parameters fulfill the conditions\footnote{This is obvious for $\tilde{\mathbf{L}}$ and $\tilde{H}$, for a proof that $\tilde{\mathbf{S}}$ is indeed unitary see Gough and James's original paper \cite{Gough2008}.} for circuit components.
Moreover, they have shown that in the case of multiple feedback loops, the result is independent of the order in which the feedback operation is applied\footnote{Note however that some care has to be taken with the indices of the feedback channels when permuting the feedback operation.}.

\appendix{Latch circuit library file}
\label{sec:latch_library}
\begin{lstlisting}[caption={Python\cite{Python} source for the pseudo-NAND latch circuit library component.}, label={lst:pseudo_nand_latch_qos}, %float=t,
language=Python]
#!/usr/bin/env python

from qhdl_component_lib.library import retrieve_component, make_namespace_string
from qnet.qos_algebra import P_sigma, qid, FB
from sympy.core.symbol import symbols

NCHANNELS = 6
GENERIC_DEFAULT_VALUES = {}

def arch_default(name_space = '', **generic_params):
    """
    Generate a symbolic LATCH expression using the provided name_space.
    Return a 2-tuple (netlist_symbolic, var_map)
    such that calling 
        >>> SLH = netlist_symbolic.substitute(var_map).evalf()
    will result in an SLH triplet, where the parameters are accessible via
        >>> SLH.S
        >>> SLH.L
        >>> SLH.H
    """
    # dictionary that stores the replacement SLH models for
    # the circuit component symbols
    var_map = {}
    
    # load symbolic component expressions as well 
    # as actual SLH model replacements for NAND1...
    NAND1, NAND1_var_map = retrieve_component('nand', 4, 
                                    make_namespace_string(name_space,'NAND1'))
    var_map.update(NAND1_var_map)
    
    #...and NAND2
    NAND2, NAND2_var_map = retrieve_component('nand', 4, 
                                    make_namespace_string(name_space,'NAND2'))
    var_map.update(NAND2_var_map)

    ########## symbolic circuit expression as computed by the parser
    # the '+' operator is overloaded to the concatenation-operation [+]
    # the '<<' operator is overloaded to the series-operation <|
    # P_sigma(s(0),s(1),...(s(n-1))) is a zero-based channel permutation object
    # FB(Q, k,l) is the feedback operation [Q]_{k->l} (with zero-based channel indices)
    # qid(n) is the n-channel identity
    netlist_symbolic = (P_sigma(0, 2, 3, 4, 5, 1)
                        << FB(
                            (
                                (qid(1) +
                                    ((qid(3) + P_sigma(2, 0, 1)) 
                                        << ((P_sigma(1, 2, 3, 0) << NAND2) + qid(2)))) 
                                << P_sigma(0, 2, 5, 6, 1, 3, 4) 
                                << ((P_sigma(0, 3, 1, 2) << NAND1) + qid(3))
                            ), 6, 1) 
                        << P_sigma(0, 4, 5, 3, 1, 2))
                        
    return netlist_symbolic, var_map

\end{lstlisting}

\FloatBarrier
\bibliography{qhdl}

\end{document}